\documentclass{PoS}
\newcommand{\qtilde}{\tilde{q}}

\title{A Lattice Calculation of Parton Distributions}

\ShortTitle{A Lattice Calculation of Parton Distributions}

\author{Constantia Alexandrou$^{ab}$, Krzysztof Cichy$^{cd}$, Kyriakos Hadjiyiannakou$^e$,\newline
Karl Jansen$^f$,  \speaker{Fernanda Steffens}$^f$%
      , Christian Wiese$^f$\\
\llap{$ˆa$} Department of Physics, University of Cyprus\\
 P.O. Box 20537, 1678 Nicosia, Cyprus\\
\llap{$ˆb$} The Cyprus Institute\\
20 Kavafi Str., Nicosia 2121, Cyprus\\
\llap{$^c$} Goethe-Universit\"{a}t Frankfurt am Main, Institute f\"{u}r Theoretische Physik\\
Max-von-Laue-Strasse 1, D-60438 Frankfurt am Main, Germany\\
\llap{$^d$} Faculty of Physics, Adam Mickiewicz Univertity\\
Umultowska 85, 61-614 Poznan\'{n}, Poland\\
\llap{$^e$} Department of Physics, The George Washington University\\
Washington, DC 20052, USA\\
\llap{$^f$} John von Neumann Institute for Computing (NIC), DESY\\
 Platanenallee 6, D-15738 Zeuthen, Germany\\
      E-mail: \email{fernanda.steffens@desy.de}}

\abstract{We present results for the $x$ dependence of the unpolarized, helicity, and transversity isovector quark distributions in the proton using lattice QCD, employing the method of quasi-distributions proposed by Ji in 2013. Compared to a previous calculation by us, the errors are reduced by a factor of about 2.5. Moreover, we present our first results for the polarized sector of the proton, which indicate an asymmetry in the proton sea in favor of the $u$ antiquarks for the case of helicity distributions, and an asymmetry in favor of the $d$ antiquarks for the case of transversity distributions.}

\FullConference{XXIV International Workshop on Deep-Inelastic Scattering and Related Subjects\\
        11-15 April, 2016\\
        DESY Hamburg, Germany}

\begin{document}

\section{Introduction}

It has been almost 50 years now since the inner structure of the proton was probed at SLAC for the first time. We have learned to a large extent on how quarks and gluons combine to give the proton its observed properties but, unfortunately, we still do not know how to calculate from first principles its partonic substructure or, for that matter, of any other hadron. The reason being that parton distribution functions (PDFs) are intrinsically non perturbative objects and, thus, to have access to them we need to solve Quantum Chromodynamics (QCD) in its non perturbative regime. The only ab initio method currently available to us to widely probe this non perturbative region is lattice QCD. Yet, these calculations of PDF's are severely limited because they are defined in the light cone, and only the lowest moments of them can be computed \cite{Alexandrou:2013cda}. In 2013, however, Ji proposed \cite{Ji:2013dva} a way to circumvent such restriction and to, finally, compute quark distributions directly on the lattice, through the use of so called quark quasi-distributions.

Quasi-distributions can be defined from a suitable choice of the Lorentz indices of the matrix element of a twist-2 operator between nucleon states, $\langle P| O^{\mu_1 \mu_2 ... \mu_n} |P \rangle = 2 \tilde{a}_n$. In the case, the choice is  $\mu_1 = \mu_2 = ... = \mu_{n} =3$, with the nucleon moving in the third direction only, $P=(P_0, 0, 0, P_3)$. Defining $\tilde{a}_{n} (P_3) = \int_{-\infty}^{+\infty} x^{n-1} \tilde{q} (x,P_3) dx$, and applying the inverse Mellin transform, one obtains for the case of unpolarized quarks:
\begin{equation}
\label{eq2}
\qtilde (x,  P_3) = \int_{-\infty}^\infty \frac{dz}{4\pi} e^{-izk_3} \langle P |
\bar{\psi}(0,z)\gamma^3 W(z) \psi(0,0) |P\rangle ,
\end{equation} 
where $k_3 = x P_3$ 
is the quark momentum in the $z$-direction, 
and $W(z) = e^{-ig \int_0^z dz^{'} A_3(z^{'})}$ is the Wilson line introduced 
to make the quark distribution gauge invariant. The relation between the quasi-quark distributions and the quark distributions at the renormalization scale $\mu_R$, $q(x, \mu_R)$, can be found, for example, in Eq. (A8) of Ref. \cite{Alexandrou:2015rja}.

In this contribution, we report on our latest effort to calculate the quark distributions using Ji's proposal for the cases of the unpolarized, helicity and transversity nonsinglet quark distributions.

\section{Lattice Setup}

We want to calculate the matrix elements of the following operators: $\gamma_3$, for the case of the unpolarized distributions (as written in Eq. (\ref{eq2})); $\gamma_3 \gamma_5$ for the case of the helicity distributions; $\gamma_3 \gamma_j$ for the case of the transversity distributions. The required matrix elements can be obtained from the ratio of suitable two-
and three-point functions. The three-point function
is constructed with the use of nucleon interpolating fields 
and a local operator:
\begin{equation}
C^{{3pt}}(t,\tau,0) = \left \langle N_{\alpha}(\vec{P},t) \mathcal O(\tau) \overline{N}_{\alpha}(\vec{P},0)\right \rangle,
\label{eq3}
\end{equation} 
where $\langle ... \rangle$ denotes the average over a sufficient number
of gauge field configurations. A nucleon field boosted with a three-momentum can be 
defined via a Fourier transformation of quark fields in position space:
\begin{equation}
N_{\alpha}(\vec{P},t) = \Gamma_{\alpha\beta} \sum_{\vec{x}}{e}^{i \vec{P} \vec{x}}\epsilon^{abc}u_{\beta}^a(x)\left( {d^b}^T(x)\mathcal C \gamma_5 u^c(x)\right),
\end{equation}
where $\mathcal C = i\gamma_0\gamma_2$ and $\Gamma_{\alpha\beta}$ is a suitable parity projector. 
Here, we will use the parity plus projector $\Gamma = \frac{1+\gamma_4}{2}$ for the unpolarized distributions,  $\Gamma = i\gamma_3 \gamma_5 \frac{1+\gamma_4}{2}$ for the helicity distributions, and  $\Gamma = i \gamma_k \frac{1+\gamma_4}{2}$ (with $k \neq j \neq 3$) for the transversity distributions given by the operator $\gamma_3 \gamma_j$.
For the case of the $\gamma_3$ operator, a vanishing momentum transfer at the operator ($Q^2=0$)
can be obtained by choosing $\mathcal O(z, \tau, Q^2=0) = \sum_{\vec{y}}\overline{\psi}(y + z)\gamma_3 W_3(y+z,y)\psi(y)$, with $y=(\vec{y},\tau)$. Similar expressions can be written for the case of the helicity and transversity operators.

Together with the two-point function, which is constructed from the nucleon
interpolating field as in Eq. (\ref{eq3}) but without the insertion of the operator, 
we can extract the desired matrix element
\begin{equation}
\frac{C^{{3pt}}(t,\tau,0;\vec{P})}{C^{{2pt}}(t,0;\vec{P})}\stackrel{0\ll \tau\ll t}{=}\frac{-iP_3}{E}h(P_3,\Delta z),
\end{equation}
with  $h(P_3,\Delta z)= \langle P |\bar{\psi}(0,z)\gamma_3 W(z) \psi(0,0) |P\rangle /2P_3$, and $E=\sqrt{(P_3)^2+M^2}$ the total energy of the
nucleon. Similarly, one can extract the matrix elements for the helicity, $\Delta h(P_3,\Delta z)$, and for the transversity, $\delta h(P_3,\Delta z)$, distributions, and for both cases the
pre-factor $\frac{-iP_3}{E}$ is absent. 

\begin{figure}
\centering
\includegraphics[scale=0.6]{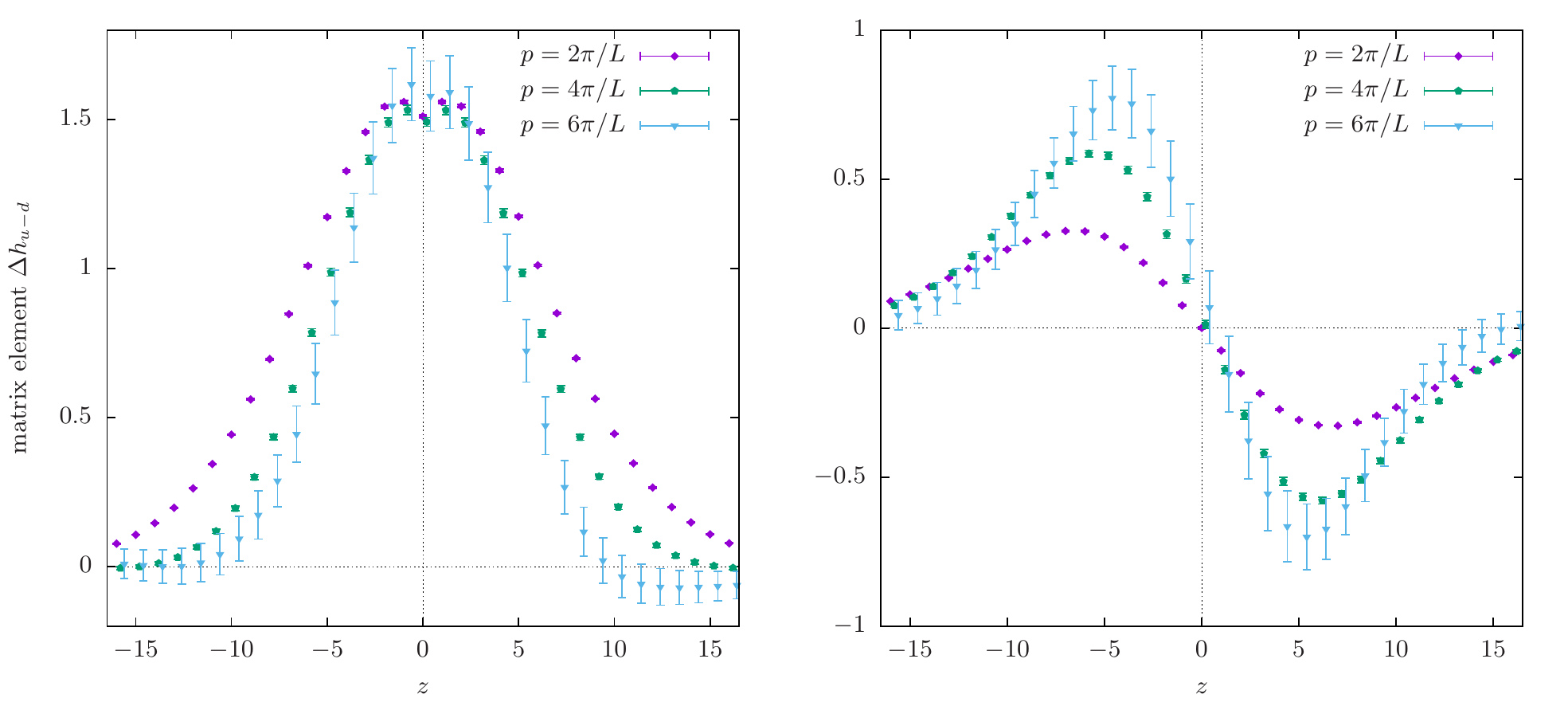}
\caption{\label{fig1} Real (left) and Imaginary (right) parts of the matrix elements for the case of the helicity operator.}
\end{figure}

We use a $32^3 \times 64$ ensemble from an ETMC (European Twisted Mass Collaboration) production ensamble \cite{Baron:2010bv}, with $N_f = 2+1+1$ flavours of maximally twisted mass fermions, with a bare coupling $\beta = 1.95$, corresponding to a lattice spacing of $a \approx 0.082$ fm. The twisted mass parameter is $a\mu = 0.0055$, which gives a pion mass of $m_{PS}\approx 370$ MeV. For the computation of the matrix elements themselves, we employ 1000 gauge configurations, each with 15 point source forward propagators and 2 stochastic propagators, resulting in total 30000 measurements, which are about 6 times more measurements than our previous results Ref. \cite{Alexandrou:2015rja}. As an example, we show in Fig. \ref{fig1} the matrix elements for the helicity operator for the first 3 lowest momenta, where 5 steps of HYP smearing \cite{Hasenfratz:2001hp} of the Wilson lines were used. We employ HYP smearing because we still do not have renormalized our operators, and it is expected that such procedure brings the value of the renormalization constants close to their tree-level values. 

\section{Results}

\begin{figure}
\centering
\includegraphics[scale=0.5]{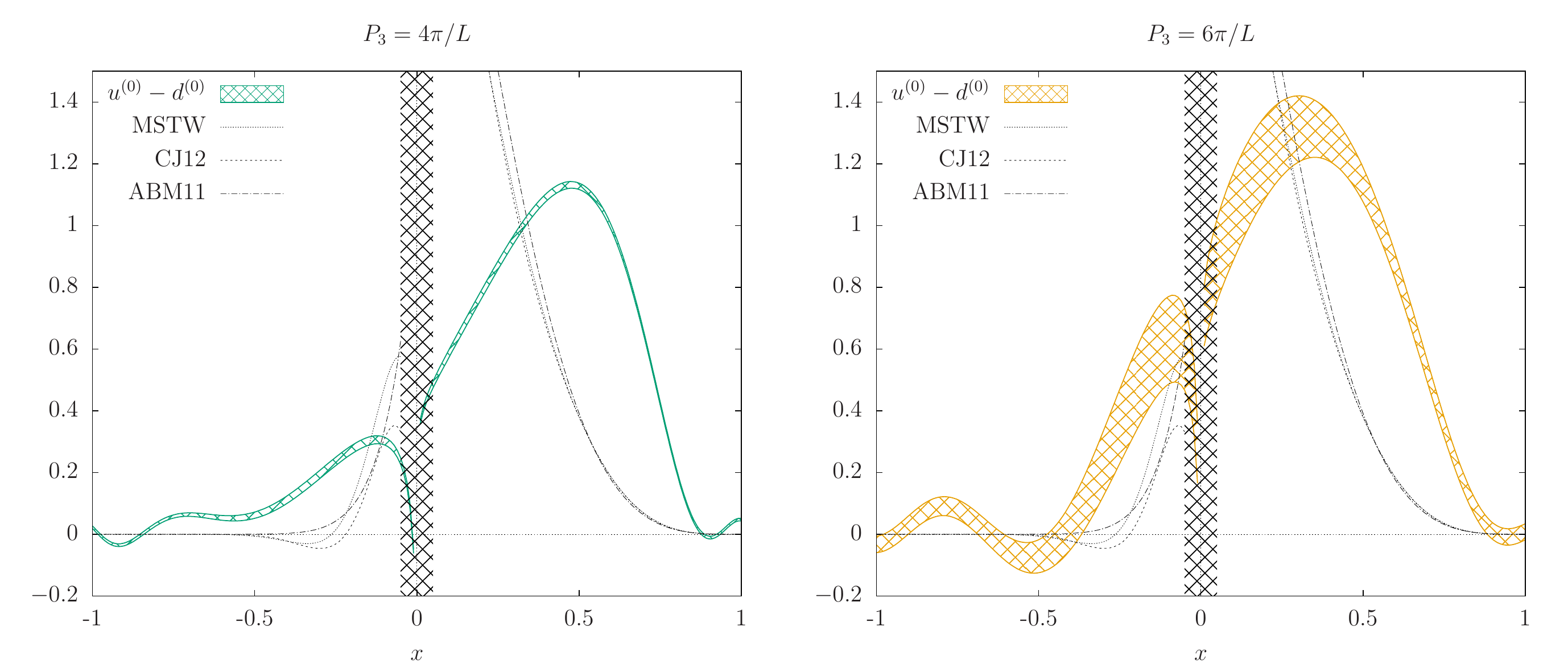}
\caption{\label{fig2} $u(x) - d(x)$ distributions at $Q^2=6.25\,$ GeV$^2$. The parameterizations for the distributions are from
MSTW \cite{Martin:2009iq}, CJ12 \cite{Owens:2012bv}, and ABM11 \cite{Alekhin:2012ig}.}
\end{figure}

Once we have the matrix elements we perform the Fourier transform as written in Eq. (\ref{eq2}) and obtain the quasi-distributions $\tilde{q}(x,P_3)$. The quasi-distributions are then corrected to take into account that the momentum of the nucleon is finite, in which case the matrix elements in Eq. (\ref{eq2}) contains a series in $M^2/P_3^2$. The relation between the corrected and non corrected quasi-distributions are given by $\tilde{q}(x,P_z) = \tilde{q}^{(0)}(\xi, P_z)/(1 + \mu \xi^2)$,
where $\xi = 2x/(1+\sqrt{1+4\mu x^2})$ is the Nachtmann variable, and superscript $(0)$ means that the nucleon mass corrections have been taken into account. In Ref. \cite{Chen:2016utp}, a different prescription has been given to take into account theses corrections. However, the difference between the two approaches are already negligible for a nucleon with momentum $6\pi/L$. 
Finally, the matching to the quark distributions $q^{(0)}(x)$ are done according to Refs. \cite{Xiong:2013bka} and \cite{Alexandrou:2015rja}. The results for the unpolarized distributions are shown in Fig. \ref{fig2}, where we plot the curves for the case of nucleon momentum $P_3 = 4\pi/L$ on the left, and for $P_3=6\pi/L$ on the right. The shaded area around $x=0$ expresses the fact that for a parton carrying momentum $xP_3$, one has in general that $x > \Lambda_{QCD}/P_3$, a restriction that is imposed from the uncertainty principle. Although we are still away from the phenomenological results, we see that there is a clear trend to approach those results as the momentum increases. Most remarkably, we see that we have a qualitative agreement with the observed $\overline{d}(x) - \overline{u}(x)$ asymmetry in the antiquark sector when we use the crossing relation $\overline{q}(x) = -\overline q(-x)$ to relate the quark distributions in the negative $x$ region to the antiquark distributions in the positive $x$ region.

\begin{figure}
\centering
\includegraphics[scale=0.5]{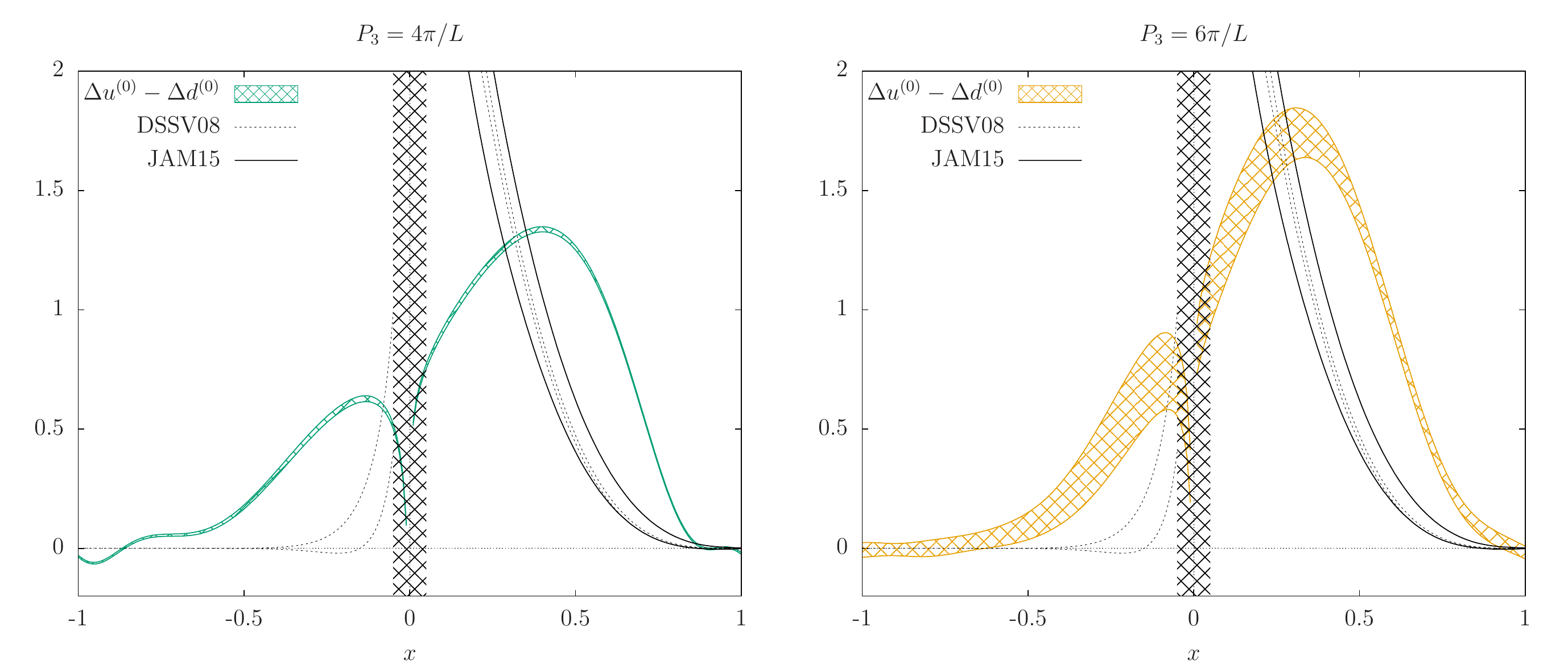}
\caption{\label{fig3} $\Delta u(x) - \Delta d(x)$ distributions at $Q^2=6.25\,$ GeV$^2$. The parameterizations for the distributions are from DSSV08 \cite{deFlorian:2009vb} and JAM15 \cite{Sato:2016tuz}}
\end{figure}

In Fig. \ref{fig3} we present the results for the helicity distributions again for 2 and 3 units of lattice nucleon momentum. The same trend observed in the unpolarized sector is seen here, that is, there is a tendency for the calculated distributions to approach the phenomenological curves as the momentum increases. For the helicity distributions, however, the crossing relation is $\Delta\overline{q}(x) = \Delta\overline q(-x)$, which implies that $\Delta\overline u > \Delta\overline d$ according to our calculation. 

\begin{figure}
\centering
\includegraphics[scale=0.5]{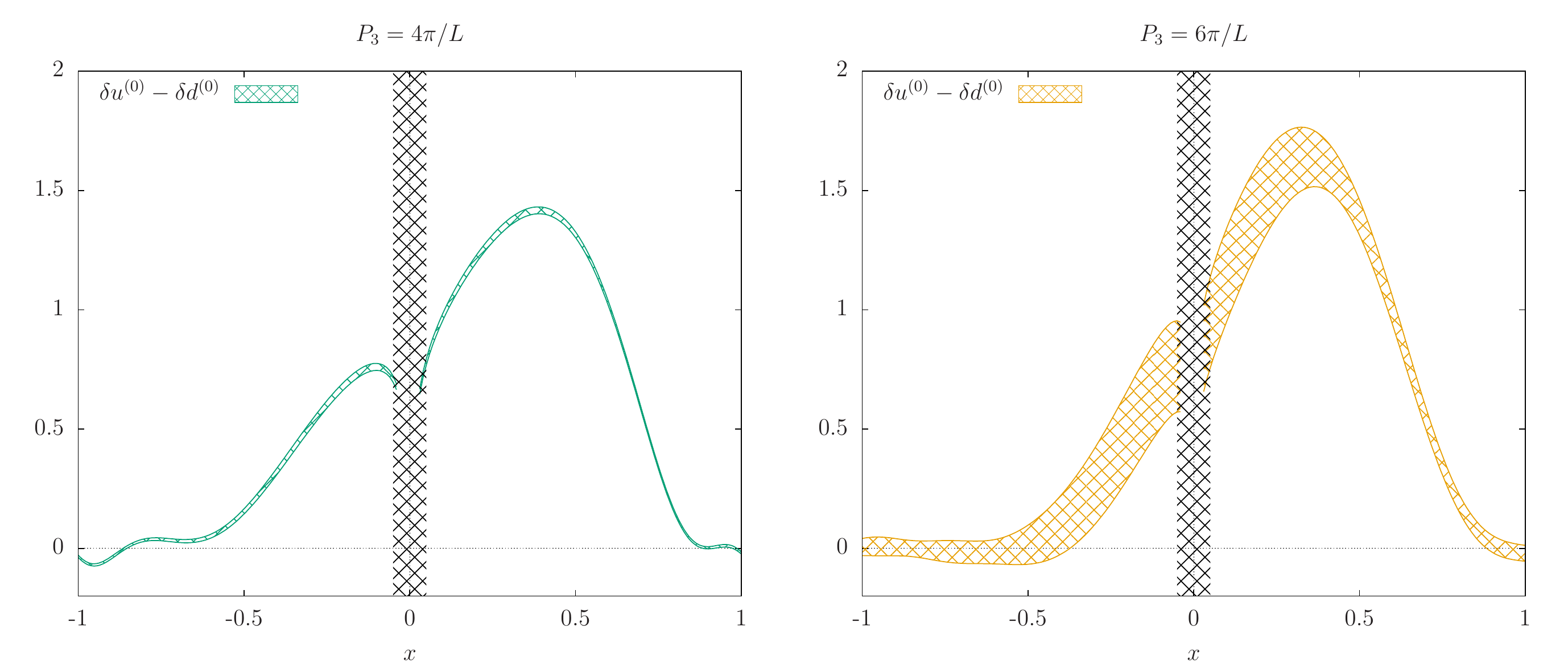}
\caption{\label{fig4} $\delta u(x) - \delta d(x)$ distributions}
\end{figure}

Finally, in Fig. \ref{fig4} we show the results for the $\delta u(x) - \delta d(x)$ transversity distributions, and in this case we do not show phenomenological parameterizations because the current ones have errors too large for the quark sector and are non existing for the antiquark sector. In any case, using the crossing relation $\delta\overline{q}(x) = -\delta\overline q(-x)$ we predict that the transversity antiquark distributions have an asymmetry similar to that of the unpolarized sector, that is, $\delta \overline d > \delta\overline u$. Similar results to ours on this matter have also been obtained in Ref. \cite{Chen:2016utp}.

\section{Perspectives}
Our lattice simulations indicate that we can reliably extract the matrix elements of the relevant operators for the complete set of the nucleon parton distributions, in the case of nonsinglet quark distributions, for a nucleon momentum up to $6\pi/L$ in lattice units. Although we have a qualitative agreement with the phenomenological curves, it is clear that we need to go to higher values of nucleon momentum if we want the lattice calculations to reproduce, and even predict, the physical results. Along with higher momentum, we also need to address the problem of renormalizability of these objects, and finally to extend these computations to the physical pion mass. These are demanding tasks that, however, have to be attacked, and we are currently addressing these problems. Our findings on these matters will be presented elsewhere in the near future.

Nevertheless, from the open points discussed above, the one related to the extension of the calculation to larger values of the boosted nucleon momentum is the one being more advanced so far. That has been possible because a new method for the smearing of the quark fields, called momentum smearing, has been proposed in Ref. \cite{Bali:2016lva}. We have implemented such a method and, from our preliminary results, we are now able to go to momentum values as high as $10\pi/L$ and $12\pi/L$. These are remarkable results because, for these values of the nucleon momentum, the quark quasi-distributions and the quark distributions, connected by a matching procedure \cite{Xiong:2013bka}, are almost equal , mainly in the large $x$ region. The same happens with the nucleon mass corrections, which become vanishingly small for large values of $P_3$. In other words, we will perform  calculations that are not, in practice, too far away from what would be calculated for a nucleon in the infinite momentum frame.
\vspace{1cm}

\hspace{-0.7cm}We are grateful to the J\"{u}lich Supercomputing Center and the DESY Zeuthen Computing Center for their computing resources and support. K.C. was supported in part by the Deutsche Forschungsgemeinschaft (DFG), project nr. CI 236/1-1 (Sachbeihilfe).

\end{document}